\documentclass[12pt,a4paper,final]{iopart}

\usepackage{iopams}  
\usepackage{graphicx}
\usepackage[breaklinks=true,colorlinks=true,linkcolor=blue,urlcolor=blue,citecolor=blue]{hyperref}
\usepackage{epstopdf}
\begin{document}
\newcommand{\be}{\begin{equation}}
\newcommand{\ee}{\end{equation}}
\newcommand{\bea}{\begin{eqnarray}}
\newcommand{\eea}{\end{eqnarray}}

\title{ Tomography on $f$-oscillators}

\author{I V Dudinets$^1$, V I Man'ko$^{1,2}$, G Marmo$^{3,4}$ and F Zaccaria$^{3,4}$ }
\address{$^1$Moscow Institute of Physics and Technology (State University),
Institutskii per. 9, Dolgoprudnii, Moscow Region 141700, Russia}
\address{$^2$ P.N. Lebedev Physical Institute, Russian Academy of Sciences, 
Leninskii Prospect 53, Moscow 119991, Russia}
\address{$^3$ Dipartimento di Fisica “E. Pancini” Universit`a di Napoli Federico II}
\address{$^4$ Istituto Nazionale di Fisica Nucleare, Sezione di Napoli,
Complesso Universitario di Monte Sant Angelo, Via Cintia I-80126 Napoli, Italy}

\ead{dudinets@phystech.edu}
\begin{abstract}
Symplectic tomographies of classical and quantum states
are shortly reviewed. The concept of nonlinear $f$-oscillators and their
properties are recalled. The tomographic probability representations of
oscillator coherent states and the problem of entanglement are then
discussed. The entanglement of even and odd $f$-coherent states is
evaluated by the linear entropy.

\end{abstract}


\submitto{\PS}
%\maketitle
\section{Introduction}
In~\cite{MankoMarmoSudarshanEtAl1997} the notion of $f$-oscillator was introduced. The notion is related to the description of specific nonlinear vibrations both in classical and quantum mechanics. Nonlinearities play an important role in different areas of physical phenomena. The specific nonlinear vibrations are relevant for problems of
quantum mechanics, for instance they can be associated with details of energy spectra of atoms and molecules or correspond to the behaviour of waves in the case of very high density of matter, which effectively produces the vibration nonlinearity from initial linear vibrations. Thus, the relevance of nonlinear vibrations in the description of physical processes at different extreme conditions is an important aspect both in classical and quantum physics. Models
of nonlinearities may be chosen to satisfy some demands, one of them would be simplicity and clear physical interpretation of the phenomena. Such models can be associated with mathematical structures like deformed algebras
or quantum groups which present the nonlinear phenomena in a form which focuses the essence of the phenomena in the structure of commutation relations of the observables. The deformed algebras or quantum groups
clarify the mathematical properties of the nonlinear models but need extra physical arguments to explain the connection of the nonlinearities in physical processes with mathematical structures used in quantum groups and
their application. One of the important and simple examples of the nonlinearity in physical process is the process by vibrations called $f$-oscillations, relevant to both domains of physics --- classical as well 
quantum. The $f$-oscillator is a generalization of the $q$-oscillator~\cite{Biedenharn1989,Macfarlane1989} discussed in literature in connection with deformations of Lie algebras (see~\cite{Daskaloyannis1999}). The $q$-oscillator is an example of a quantum integrable system~\cite{clemente2009towards}. Some aspects of the $q$-oscillator were cosidered in~\cite{MankoMarmoSolimenoEtAl1993}. A suggestion to describe the possible nonlinearity of classical electrodynamics by the $q$-oscillators is proposed in~\cite{man1993correlation}. The $f$-oscillator is the nonlinear oscillator with the specific dependence of the frequency of the vibrations on the amplitude. The eigenvalue problems for the Hamiltonians describing different kinds of $f$-nonlinearities were studied in detail by Dodonov et al.~\cite{DodonovSpectrum}. In these days the nonlinear oscillator attracts attention in connection with
the study of nonclassical photon states in quantum optics~\cite{FaghihiTavassoly2013,FaghihiTavassolyHooshmandasl2013,AliGazeauHeller2008,MatosFilhoVogel1996}. The possibility to take into account the $f$-nonlinearity of vibrations in some cases of laser light radiation and its
coherence was studied recently by Kilin et al.~\cite{KilinMikhalychev2012}. 
Nonclassical properties such as squeezing, antibunching, sub-poissonian statistics of nonlinear coherent states were considered in~\cite{roy2000new,safaeian2011deformed}.
Theoretical schemes for realization of the nonlinear coherent states have been proposed based on the
trapped ion~\cite{MatosFilhoVogel1996}, 
an optomechanical microcavity~\cite{Yan16},
single atom laser~\cite{kilin2012single},
exciton dynamic in a quantum dot~\cite{harouni2009q},
spatial confinement of a harmonic oscillator~\cite{harouni2008spatial},
 a particle in a finite range trap~\cite{darareh2010nonclassical,harouni2011single} or 
the oscillation and vibration of the graphene membrane~\cite{harouni2016preparation}.
Coherent states are also called the most classical states of light  because they minimize the
Heisenberg uncertainty relations~\cite{heisenberg1927anschaulichen}. Superposition of coherent states possess  statistical properties different
from the coherent states. Even or odd  superpositions of coherent states were considered in~\cite{DodonovMalkinManKo1974}. 
Superposition of coherent states related to a finite group ('crystallised Schrodinger cats') 
is studied in~\cite{castanos1995crystallized}. Extension of this superposition to the case of nonlinear coherent states is considered in~\cite{abbasi2017four}.
Recently, method for producing of the superposition of nonlinear coherent states (as well as entangled nonlinear coherent states) is suggested in~\cite{karimi2015production,perez2016generalized}. 
Entangled states such as entangled coherent states are of great importance due to various application in quantum information theory.
In this connection, different experimental schemes were proposed to generate entangled coherent states. 	
Producing of entangled nonlinear coherent states and their properties were considered in ~\cite{karimi2016production,afshar2016nonclassical,honarasa2016entanglement}.
An oscillator with spatially varying mass can be considered as the $f$-oscillator with  the special type of deformation function ~\cite{amir2016barut}.  A two-dimensional harmonic oscillator algebra on a sphere can be considered as a deformed one-dimensional harmonic oscillator algebra~\cite{mahdifar2006geometric}. A recent proposal to use $f$-oscillators in the description of the Hydrogen atom has been published recently~\cite{ClementeMarmo}.
Different construction of nonlinear coherent states is proposed in~\cite{penson1999new}. Also $f$-deformation algebra arises in connection with quantizing dynamically equivalent Hamiltonian structures~\cite{stichel2000dynamical}.  
The conventional coherent states~\cite{Klauder1963,Glauber1963,Sudarshan1963} correspond
to maximally classical electromagnetic field vibrations which minimize
uncertainty relations~\cite{heisenberg1927anschaulichen,SudarshanChiuBhamathi1995,DodonovKurmyshevManko1980}. The nonlinear vibrations of the $f$-oscillator make the quantum features of photons more pronounced like the existence of bounds in the product of photon quadrature variances which is larger than usual quantum limit due to
the nonlinearity of the vibrations. The Schr\"odinger-Robertson~\cite{Schroedinger1930,Robertson1930} uncertainty relation with the bound for the product of positions and momentum dependent on their covariance was studied in the context of the influence of the nonlinearity contributions in the work by Bastos et al.~\cite{BastosBertolamiDiasEtAl2012}. An assumption that quantum geometry effects lead to random fluctuations of the Planck constant thus changing the uncertainty relation is given in~\cite{mangano2015inconstant}. A suggestion to check the uncertainty relation in experiments with homodyne photon states detection was presented in~\cite{ManKoMarmoPorzioEtAl2011}. In the homodyne detection experiments the optical tomogram of photon states is measured~\cite{ManKoMarmoPorzioEtAl2011,SmitheyBeckRaymerEtAl1993,BelliniCoelhoFilippovEtAl2012}. Such experiments influenced the development of tomographic probability representation of
quantum mechanics and quantum optics~\cite{ManciniMankoTombesi1996,IbortManKoMarmoEtAl2009}.
Recently~\cite{MankoMarmoZaccaria2010} the discussion of some tomographic aspects of the $f$-oscillator was given for the symplectic tomography scheme. It is known that there exist also the schemes of the center of mass tomographic representation~\cite{ArkhipovPRA} and the photon number tomographic probability representation~\cite{ManciniTombesiManko1997}. In all such
representations quantum states are identified with fair positive probability
distributions of measurable observables and the density operators as well as
the variances, covariances and higher moments of physical observables can be
explicitly expressed in terms of the tomographic probability representation.
In view of these the tomograms are, in facts, alternative to density
operators or wave functions. The connection of the $f$-oscillators with analogs
of Weyl systems was studied in~\cite{AnielloMankoMarmoEtAl2000}. Some tomographic aspects of such oscillators were studied in~\cite{man2008tomographic}. The aim of the present work
is to study the states of $f$-oscillators in the frame of tomographic
probability representations, in particular, extending the consideration and
results of the work~\cite{MankoMarmoZaccaria2010} to the case of several
modes and also to the case of the photon number tomographies. We also consider how some specific choices of the nonlinearity function $f$ may introduce entanglement among different modes.

The paper is organized as follows. 

In Sec.~2 the review of the one-mode $f$-oscillator properties is presented. In Sec.~3 tomographic probability representations including symplectic, optical and photon number cases are considered. Explicit formulae for symplectic and photon number tomograms of the $f$-oscillator are given in Sec.~4. In Sec.~5 the photon number tomogram of the Fock states is used to construct new inequalities for the associated Laguerre polynomials. Entanglement of two-mode nonlinear coherent states is studied in Sec.~6 and its superposition in Sec.~7. In Sec.~8 the Schr\"odinger-Robertson uncertainty relation for deformed position and momentum is considered. Conclusions and perspectives are summarized in Sec.~9.

\section{Classical and quantum $f$-oscillators for one-mode fields}
\hspace*{1.5em}
Dynamics of a classical linear oscillator vibrating with unit frequency is determined by the equation of motion, which in terms of the complex amplitude of the oscillator has the form
\be
\frac{\rmd \alpha}{\rmd t}=-\rmi\alpha,
\ee
where $\alpha=(q+\rmi p)/\sqrt{2}$ with $q$ and $p$ being position and momentum of the classical oscillator.
A solution to the latter equation reads
\be
\alpha (t)=\alpha (0) \rme ^{-\rmi t},
\ee
where $\alpha(0)$ is any initial complex amplitude.
It is obvious that the energy of the oscillator is the integral of motion 
\be\label{energy}
E=\alpha (t)\alpha^*(t)=\alpha (0)\alpha^*(0).
\ee
A direct generalisation of the linear harmonic oscillator to the nonlinear one is carried out by the replacement of the complex amplitude $\alpha$ with the deformed one $\alpha_f$
\be
\alpha_f=\alpha f(\alpha \alpha^*),
\ee
where the real function $f$ governs the nonlinearity of vibrations. To this end, consider the $f$-oscillator, which is  the nonlinear oscillator described by the Hamiltonian 
\be
H_f=\alpha_f \alpha_f^*.
\ee
The equation of motion of the nonlinear oscillator with the Hamiltonian $H_f$ reads
\be\label{motion_f}
\frac{\rmd \alpha}{\rmd t}=-\rmi \omega (\alpha \alpha^*)\alpha,
\ee
where $\omega (E)=\partial (E f^2(E))/\partial E$ is the frequency of vibrations and $E$ is given by~(\ref{energy}).
Since $E$ is an integral of motion for both the linear and nonlinear oscillators, one can easily obtain the solution to~(\ref{motion_f})
\be
\alpha (t)=\alpha (0) \rme^{-\rmi \omega (\alpha(0)\alpha^*(0))t}.
\ee
To sum up, the $f$-oscillator is a nonlinear system with a particular type of nonlinearity, namely the frequency of vibrations depends on its amplitude. 

For a quantum harmonic oscillator the position $q$ and momentum $p$ of the classical one are replaced by the corresponding operators $\hat{q}$ and $\hat{p}$, while the complex amplitudes $\alpha$ and $\alpha^*$ are replaced by the annihilation and creation operators $\hat{a}$ and $\hat{a}^{\dagger }$ obeying the commutation relation
\be  \label{Eq:boson_comm_rel}
\left[ \hat{a},\hat{a}^{\dagger }\right] =1. 
\ee
One can perform a quantization procedure analogous to the one of the linear oscillator for the classical $f$-oscillator by selecting an ordering and by replacing  the complex amplitudes $\alpha_f$ and $\alpha^*_f$ with the deformed annihilation and creation operators
\be
\hat{A}=\hat{a}f\left( \hat{a}^{\dagger }\hat{a}\right), \quad
\hat{A}^{\dagger }=f\left( \hat{a}^{\dagger }\hat{a}\right) \hat{a}
^{\dagger }
\ee
with $f$ being a real operator valued function of the number operator $\hat{a}^{\dagger }\hat{a}$.
By definition, the $f$-oscillator is characterized by the Hamiltonian of the form
\be
\hat{H}_f=(\hat{A}\hat{A}^{\dagger}+\hat{A}^{\dagger}\hat{A})/2.
\ee
The commutation relation between the deformed operators is
\be\label{commut_rel}
\left[ \hat{A},\hat{A}^{\dagger}\right]=(\hat{a}^{\dagger}\hat{a}+1)f^2(\hat{a}^{\dagger}\hat{a}+1)-\hat{a}^{\dagger}\hat{a}f^2(\hat{a}^{\dagger}\hat{a}).
\ee

The $f$-coherent states is defined as an eigenstate of the deformed annihilation operator
\be
\hat{A}|\alpha,f\rangle = \alpha |\alpha,f\rangle
\ee
with $\alpha$ being a complex number. In terms of the Fock states the state represented  in a compact form by
\be\label{composition_f}
|\alpha,f\rangle=N_f (\alpha)\sum_{n=0}^{\infty}\frac{\alpha^n}{\sqrt{n!} f(n)!}|n\rangle.
\ee
where we denote $f(n)!=f(0)f(1)f(2)\ldots f(n)$. The normalization constant reads 
$N_f(\alpha) = \left [\sum_{n=0}^{\infty}\frac{|\alpha|^{2n}}{n! [f(n)!]^2}\right]^{-1/2} $.
For the choice $f\left( \hat{a}^{\dagger }\hat{a}\right)=1$ the $f$-coherent state give back the usual coherent state, 
i.e. $|\alpha,f=1\rangle=|\alpha\rangle$.

\section{Tomographic probability distributions}

In this section we briefly review special types of maps of density operators onto probability distribution functions.
The symplectic tomographic probability representation of a quantum state
with the density operator $\hat{\rho}$ is given in terms of the mean value of the
dequantizer operator~\cite{MankoMarmoSudarshanEtAl2002}
\be
M(X,\mu ,\nu )=\langle\delta \left( X\hat{1}-\mu \,\hat{q}-\nu\, \hat{p}\right) \rangle=\Tr [
\hat{\rho}\,\delta \left( X\hat{1}-\mu \,\hat{q}-\nu \,\hat{p}\right) ].
\label{symplectic}
\ee
Here $X$ is a random homodyne quadrature, $\mu $ and $\nu $ are real
parameters and $\hat{1}$, $\hat{q}$, $\hat{p}$ are the identity, momentum and position operators, respectively.
The optical tomogram is defined in view of the symplectic tomogram as follows
\begin{equation}
w\left( X,\theta \right) =\langle\delta \left( X\hat{1}-\hat{q}\cos \theta -\hat{p}
\sin \theta \right)\rangle.
\end{equation}

It is worth to point out that both the symplectic tomogram $M(X,\mu ,\nu )$
and the optical tomogram $w\left( X,\theta \right) $ are conditional
probability distributions of the photon homodyne quadrature $X$ provided the
real parameters $\mu$ and $\nu $ are fixed (for the symplectic tomogram) and the
local oscillator phase $0\leq \theta \leq 2\pi $ is fixed (for the optical tomogram)~\cite{man2012tomographic}. 
The optical and symplectic tomograms are nonnegative normalized functions 
\be
\int  M(X,\mu ,\nu )\,dX =1\ , \quad
\int w\left( X,\theta \right)\,dX =1.  
\ee

The important feature of the optical tomogram is connected with the fact
that it is directly measurable in experiments on homodyne quadrature
detection of the photon states~\cite{SmitheyBeckRaymerEtAl1993}.

In view of this fact the measurement can be used to obtain all
the characteristic of quantum objects like any observables in terms of its
means, dispersions, highest moments, etc. using only the tomograms and
avoiding the use of wave functions or density operators.
Given a state with the density operator $\hat{\rho}$ higher moments of the homodyne
quadrature $X$ are expressed in terms of the optical tomogram as 
\be
\langle X^{k}\rangle=\Tr\left[ \hat{\rho}\left( \hat{q}\cos \theta +\hat{p}\sin
\theta \right)^k \right] =\int X^{k}\,w\left( X,\theta \right)dX .
\ee

For example, the above formula provides the position higher moments $\langle\hat{q}^{k}\rangle=\int X^{k}\,w\left( X,\theta=0 \right)dX $ or the momentum statistics $\langle\hat{p}^{k}\rangle=\int X^{k}\,w\left( X,\theta=\frac{\pi}{2} \right)dX $. The latter fact gives rise to the possibility to check the uncertainty relations with the help of the measurable optical tomogram. There exist inversion relations expressing the density operator $\hat{\rho}$ in terms of either the optical, symplectic or photon number tomograms (see e.g.~\cite{IbortManKoMarmoEtAl2009}).

The photon number tomogram is defined as the diagonal matrix element of the displaced density operator
\be\label{phot_numb_tom}
\mathcal{W}\left( n,\alpha \right) =\langle n| \mathit{\hat{D}}\left( \alpha \right)\hat{\rho}\mathit{\hat{D}}
^{\dagger }\left( \alpha \right)  
|n\rangle.
\ee
In this definition, $n$ is the number of photons, $\alpha = (q+ip)/\sqrt{2}$ is a complex field amplitude
and $\mathit{\hat{D}}\left( \alpha \right) $ is the displacement operator
$ \mathit{\hat{D}}\left( \alpha \right) =\exp \left( \alpha \,\hat{a}^{\dagger
}-\alpha ^{\ast }\,\hat{a}\right).$
The creation $\hat{a}^{\dagger }$ and annihilation $\hat{a}$ operators of
photons satisfy the basic commutation relations~(\ref{Eq:boson_comm_rel}).\
The tomogram $\mathcal{W}\left( n,\alpha \right) $ is the conditional
probability distribution of photons provided the complex field amplitude $
\alpha $ is fixed~\cite{man2012tomographic}. The photon number tomogram is a normalized function over the discrete variable $n$ and the continuous complex variable $\alpha$ 
\be
\sum\limits_{n=0}^{\infty }\mathcal{W}\left( n,\alpha \right) =1, \quad 
\int \mathcal{W}\left( n,\alpha \right)\frac{d^2 \alpha}{\pi}=1.
\ee
For a pure state vector $|\psi\rangle$ the transform~(\ref{phot_numb_tom}) yields
\be
\mathcal{W}\left( n,\alpha \right) =\left | \langle n|\mathit{\hat{D}}\left( \alpha \right)|\psi\rangle \right |^{2}.
\ee
The Husimi function of a density operator $\hat{\rho}$ is defined as $Q(\alpha)=\langle \alpha |\hat{\rho}|\alpha\rangle$, can be simply obtained from the photon number tomogram by the following formula
\be\label{Husimi_rule}
Q(\alpha)=\mathcal{W}(n=0,-\alpha).
\ee
The photon number tomogram provides the probability of finding $n$ photons in the state $\hat{\rho}$
\be
P(n)\equiv  \langle n|\hat{\rho}|n\rangle=\mathcal{W}(n,\alpha =0).
\ee

The tomographic probability distributions provide all the information on
quadrature states. For example, the photon number tomogram yields all the
statistical properties of photons. Higher moments of the photon
number $\hat{n}=\hat{a}^{\dagger}\hat{a}$ in a quantum state with the density operator $\hat{\rho}$ read 
\be
\langle \hat{n}^{k}\rangle=\Tr\left[ \hat{\rho}\left( \hat{a}^{\dagger }\hat{a}\right)
^{k}\right] =\sum\limits_{n=0}^{\infty }\mathcal{W}\left( n,0\right) n^{k},
\ee
for an arbitrary nonnegative integer $k$.

\section{Tomograms for $f$-oscillator}
In this section we introduce examples of tomographic probability distributions. Calculating the optical tomogram of a pure state found in a state $\psi_0(x)$ is an intractable task because it requires performing the fractional Fourier transform of the wave function under consideration~\cite{man1999non}. 
However, in some cases this may be circumvented by using the fact that the optical tomogram formally equals to the modulus squared of the wave function $\psi(X,t)$ taken at time $t=\theta$, which is the solution of the Shr\"oedinger evolution equation for a harmonic oscillator potential provided the wave function at initial time equals $\psi_0(x)$~\cite{de2006new}. To this end, we consider a wide class of quantum states for which one can obtain the wave function $\psi(X,t)$. We begin with a state, which expressed in the Fock number states in the most generic case
\be \label{gener}
 |\psi\rangle = \sum_{n=0}^{\infty}c_n|n\rangle.
\ee
The state is assumed to be normalized, i.e. $\sum_{n=0}^{\infty} |c_n|^2=1$.
The symplectic tomogram of this state can be directly calculated from definition~(\ref{symplectic}). However, here we provide a simpler method. The temporal evolution of the state for the harmonic oscillator potential is
\be \label{evol}
 |\psi,t\rangle = \sum_{n=0}^{\infty}c_n \rme^{-\rmi E_n t}|n\rangle.
\ee
Here $E_n$ corresponds to the energy spectrum of the harmonic oscillator with unity frequency and the Planck constant, i.e. $E_n=n+\frac{1}{2}$. The wave function~(\ref{evol}) in position representation reads
\be\label{evol2}
\psi (X,t)=  \sum_{n=0}^{\infty}c_n \rme^{-\rmi(n+\frac{1}{2})\, t} \psi _n(X),
\ee 
where the initial wave function $\psi _n(X)=\pi^{-1/4}\,2^{-n/2}\,(n!)^{-1/2}\,\exp{(-X^2/2)}\,H_n(X)$ is  expressed in terms of the Hermite polynomial  $H_n$ of degree $n$.  
The modulus squared of~(\ref{evol2}) taken at time $t=\theta$ provides the optical tomogram
\be\label{opt_gener}
w(X,\theta)=\left|\psi (X,t=\theta) \right|^2=\frac{\exp{\left(-X^2\right)}}{\sqrt{\pi}}\,\left | 
\sum_{n=0}^{\infty}\frac{c_n}{\sqrt{n!}}\frac{\rme^{-\rmi n \theta}}{2^{n/2}} 
H_n\left ( X \right ) \right |^2.
\ee
Let us note that the expression at the right hand side of~(\ref{opt_gener}) is periodic function over $\theta$ with period $2\pi$. It can be readily seen that $w(X,\theta)$~(\ref{opt_gener}) is a nonnegative function. For the given optical tomogram one can reconstruct the symplectic tomogram by the following rule
\be
 M(X,\mu ,\nu )=\left(\mu^2+\nu^2\right)^{-1/2}\,w \left(\frac{X}{\sqrt{\mu^2+\nu^2}},\theta \right)
\ee   
provided $\cos\theta=\mu/\sqrt{\mu^2+\nu^2}$ and $\sin\theta=\nu/\sqrt{\mu^2+\nu^2}$. The result reads
\be\label{sym_tom_gener}
 \fl M(X,\mu ,\nu )=\frac{\exp{\left(-\frac{X^2}{\mu^2+\nu^2}\right)}}{\sqrt{\pi(\mu^2+\nu^2)}}\,\left | 
\sum_{n=0}^{\infty}\frac{c_n}{\sqrt{n!}}\left (\frac{ \mu-i\nu}{\sqrt{2}\sqrt{\mu^2+\nu^2}} \right )^n
H_n\left ( \frac{X}{\sqrt{\mu^2+\nu^2}} \right ) \right |^2.
\ee

It is convenient to apply the latter formula only for states which can be simply decomposed into the Fock basis states. For example, the symplectic tomogram for the bound state of a particle moving in the Dirac delta potential, which expressed in terms of the Error function~\cite{dudinets2013bound} can be easily obtained from its definition rather than from~(\ref{sym_tom_gener}).

As an example, let us consider the $f$-coherent state $|\alpha,f_{\lambda}\rangle $ corresponding to the nonlinear function \begin{equation}\label{KerrDeform}
f_{\lambda}(\hat{n})=\lambda^{-1/2} \sqrt{\hat{n}-1+\lambda}.
\end{equation}
There exist several examples where this state might be useful. For example, the Kerr effect admits description by this type of $f$-oscillators with $\lambda^{-1}$ being  the crystal
nonlinear susceptibility~\cite{osborn2009moyal} or the Hamiltonian of the $f$-oscillator has eigenvalues identical to the energy spectrum of the trigonometric P\"oschl-Teller potential~\cite{antoine2001temporally}. The expression of $|\alpha,f_{\lambda}\rangle$ in the number state basis is
\be 
|\alpha,f_{\lambda}\rangle=\left[{}_0F_1\left(\lambda,\lambda |\alpha|^2 \right)\right]^{-1/2}\sum_{n=0}^{\infty}\alpha^n\sqrt{\frac{\lambda^n \Gamma (\lambda)}{n!\, \Gamma (\lambda+n)}} |n\rangle.
\ee
Here ${}_0F_1 (a,z)=\sum_{n=0}^{\infty}z^n \Gamma (a)/\left( n! \,\Gamma (a+n)\right)$ is the confluent hypergeometric limit function. The symplectic tomogram of this state is 
\begin{eqnarray}
 M_{\lambda}(X,\mu ,\nu )=\frac{1}{{}_0F_1\left(\lambda,\lambda |\alpha|^2 \right)}\frac{\exp{\left(-\frac{X^2}{\mu^2+\nu^2}\right)}}{\sqrt{\pi(\mu^2+\nu^2)}}\nonumber\\
\times \left | 
\sum_{n=0}^{\infty}\frac{\alpha^n}{n!}\sqrt{\frac{\lambda^n \Gamma (\lambda)}{ \Gamma (\lambda+n)}} \left (\frac{ \mu-i\nu}{\sqrt{2}\sqrt{\mu^2+\nu^2}} \right )^n
H_n\left ( \frac{X}{\sqrt{\mu^2+\nu^2}} \right ) \right |^2.
\end{eqnarray}

 The photon number tomogram~(\ref{phot_numb_tom}) of the state~(\ref{gener}) is expressed in terms of matrix elements of the displacement operator and is given through the associated Laguerre  polynomials 
\begin{eqnarray}
\label{phot_gener}
\mathcal{W}\left( n,\alpha \right)=\exp{\left(-|\alpha|^2\right)} \Bigg| \sum_{m=0}^{n}c_m \sqrt{\frac{m!}{n!}} \,\alpha^{n-m}\,L^{(n-m)}_m(|\alpha|^2)\nonumber\\
+\sum_{m=n+1}^{\infty}c_m \sqrt{\frac{n!}{m!}} \,(-\alpha^*)^{m-n}\,L^{(m-n)}_n(|\alpha|^2) \Bigg |^2.
\end{eqnarray}
The Husimi function can be obtained from~(\ref{Husimi_rule}) using the known property of the Laguerre polynomials~\cite{gradshteyn2014table}
$L^{(m)}_0(x)=1$
\begin{equation}
Q(\alpha)=\exp(-|\alpha|^2)\left|\sum_{m=0}^{\infty}\frac{c_m (\alpha^*)^m}{\sqrt{m!}} \right |^2.
\end{equation}
Nevertheless, the latter formula can be obtained using the definition of the Husimi function $Q(\alpha)=
\left| \langle \alpha |\psi \rangle \right|^2$ and the series expansion~(\ref{gener}).  
For $\alpha \rightarrow 0$ we get the photon probability function $P(n)=|c_n|^2$. 

The photon number tomogram of the Fock space $|m\rangle$ is expressed in  terms of the associated Laguerre  polynomials as follows
\be\label{pm_def}
\fl \mathcal{W}_{m}\left( n,\alpha \right)=\cases{\frac{n!}{m!}|\alpha|^{2m-2n}\exp \left( -|\alpha
|^{2}\right) \left [L_{n}^{(m-n)}\left( |\alpha |^{2}\right)   \right ]^2,&for $m\geq n$\\
\frac{m!}{n!}|\alpha|^{2n-2m}\exp \left( -|\alpha
|^{2}\right) \left [L_{m}^{(n-m)}\left( |\alpha |^{2}\right)   \right ]^2,&for $m\leq n$.\\}
\ee
It is worth noting that the photon number tomogram~(\ref{pm_def}) is symmetric with respect to $n$ and $m$. This means that if one interchanges $n$ and $m$, the function values do not change. 
The tomogram~(\ref{pm_def}) provides the probability distribution $p_m$, which satisfies entropic inequalities.   

\section{Entropic inequalities for associated Laguerre polynomials}

In this section we make use of the approach developed in~\cite{man2015properties} to obtain inequalities for the associated Laguerre polynomials. Let us consider an arbitrary set of nonnegative numbers $p_m$, $m=0,1,\ldots $ with the sum equal to one $\sum_{m=0}^{\infty} p_m=1$. The set of numbers  can be interpreted as a probability distribution of a classical system with one random variable and provides the Shannon entropy
\be \label{Shannon}
H_p=-\sum_{m=0}^{\infty} p_m \ln p_m.
\ee
Following~\cite{man2015properties} let us introduce the invertible map of the nonnegative numbers $p_m$ onto nonnegative numbers $\mathcal{P}_{jl}$
\be\label{map2}
p_{2j+l}\leftrightarrow \mathcal{P}_{jl},
\ee
where $j$ is a nonnegative integer number and $l$ takes two values $0$ and $1$. Since the numbers $\mathcal{P}_{jl}$ can be interpreted as joint probability distribution of a bipartite classical system with two random variables, one can introduce other nonnegative numbers corresponding to marginal probability distributions 
\be 
\Pi_j=\sum_{l}\mathcal{P}_{jl},\quad \pi_l=\sum_{j}\mathcal{P}_{jl},
\ee
which provide the Shannon entropies
\be\label{marg_entr}
H_{\Pi}=-\sum_{j} \Pi_j \ln \Pi_j,\quad H_{\pi}=-\sum_{l} \pi_l \ln \pi_l.
\ee
The entropies of a bipartite system and its subsystems are known to satisfy the entropic inequalities called the subadditivity condition~\cite{lieb1973proof}  
\be \label{subaddit}
H_{\Pi}+H_{\pi}\geq H_{p}.
\ee
The latter inequality in terms of the probabilities $p_m$ has the form
\begin{eqnarray}
\label{ineqentr}
 -\sum_{m=0}^{\infty} \left (p_{2m}+p_{2m+1}  \right ) \ln \left ({p_{2m}+p_{2m+1}}  \right )
- \left (\sum_{m=0}^{\infty}p_{2m}  \right ) \ln \left (\sum_{m=0}^{\infty}p_{2m}  \right ) \nonumber \\
- \left (\sum_{m=0}^{\infty}p_{2m+1}  \right ) \ln \left (\sum_{m=0}^{\infty}p_{2m+1}  \right )\geq  
 -\sum_{m=0}^{\infty} p_m \ln p_m. 
\end{eqnarray}
The Shannon information (also called the mutual information) is defined as 
\be \label{inform}
I=H_{\Pi}+H_{\pi}-H_p.
\ee
Obviously, the Shannon information is a nonnegative function. The Shannon information is used to measure correlations between subsystems. The point is that the photon number tomogram ~(\ref{pm_def}) is nonnegative with unity sum over $m$ for all values $n$ and $\alpha$. Hence, $\mathcal{W}_{m}\left( n,\alpha \right)$ can be considered as probabilities
\be \label{pm_def2}
p_m (n,\alpha)=\mathcal{W}_{m}\left( n,\alpha \right),
\ee
where $\mathcal{W}_{m}\left( n,\alpha \right)$ is given by~(\ref{pm_def}). 
Substituting the latter expression for $p_m$ into~(\ref{ineqentr}), we obtain the new inequality for   
the associated Laguerre polynomials
\begin{eqnarray}
\label{Laguerre_ineq}
-\sum_{m=0}^{\infty} \left (\lambda_{2m}(n,x)+\lambda_{2m+1}(n,x)  \right ) \ln \left ({\lambda_{2m}(n,x)+\lambda_{2m+1}(n,x)}  \right )\nonumber\\ \fl
- \left (\sum_{m=0}^{\infty}\lambda_{2m}(n,x)  \right ) \ln \left (\sum_{m=0}^{\infty}\lambda_{2m}(n,x)  \right ) 
- \left (\sum_{m=0}^{\infty}\lambda_{2m+1}(n,x)  \right ) \ln \left (\sum_{m=0}^{\infty}\lambda_{2m+1}(n,x)  \right )
 \nonumber \\+\sum_{m=0}^{\infty} \lambda_m(n,x) \ln \lambda_m(n,x)+x\,\rme^x \geq  0, 
\end{eqnarray}
where the notations $x=|\alpha|^2\geq 0$ and 
\begin{equation}\label{lambda}
\lambda_m(n,x)=\cases{\frac{n!}{m!}x^{m-n}  \left [L_{n}^{(m-n)} (x)   \right ]^2,&for $m\geq n$\\
\frac{m!}{n!}x^{n-m} \left [L_{m}^{(n-m)}(x)   \right ]^2,&for $m\leq n$\\}
\end{equation} are used.

\begin{figure}[ht]
\center{ \includegraphics[width=10.6cm]{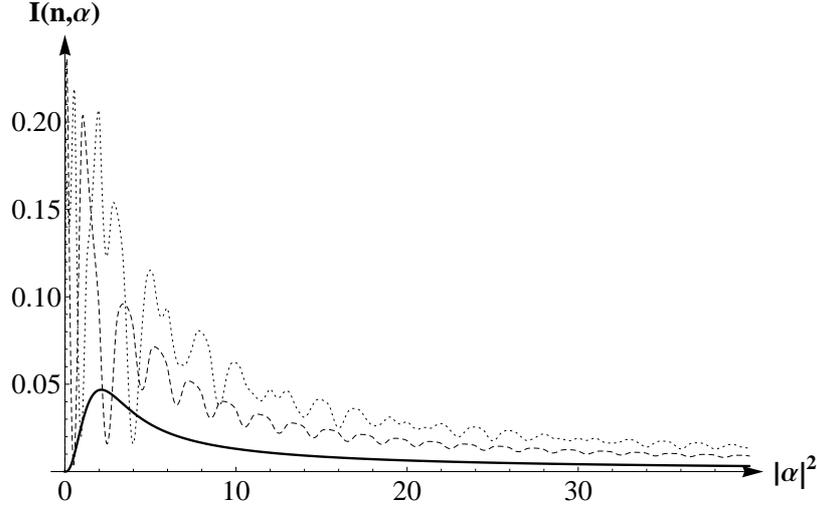}}
\caption{The Shannon information $I(n,\alpha)$~(\ref{inform}) for probabilities~(\ref{pm_def2}) for $n=0$ (solid line), $n=1$ (dashed line), $n=2$ (dotted line) and various 
values of $|\alpha|^2$}
\label{fig:inform}
\end{figure}

Thus, we have interpreted the set of nonnegative numbers with unit sum as the probability distributions  of some artificial bipartite system. This fact gives the possibility to apply the well-known entropic inequality to the bipartite system and its subsystems. The relation obtained reflects the presence of correlations between the subsystems.  To measure correlation we have used  the Shannon information. The Shannon information $I(n,\alpha)$~(\ref{inform}) for probabilities~(\ref{pm_def2}) is shown in Fig.~\ref{fig:inform}. One can see that for large values of $|\alpha|^2$
($|\alpha|^2$ is related to the energy of the field) the Shannon information tends to zero and the subsystems (the first one contains odd photon numbers and the second one contains even photon numbers) become independent. On the contrary, for small values of $|\alpha|^2$ there exists a correlation between the subsystems. Also, on the whole for fixed $|\alpha|^2$ the correlation increases when the number of photon $n$ increases.  The information oscillates for $n\geq 1$ and is smooth for $n=0$, such behavior is associated with the zeroes of the Laguerre polynomials.

There are many other possibilities to map the numbers $p_m$ onto $\mathcal{P}_{jl}$ and hence obtain new inequalities for the associated Laguerre polynomials. In~(\ref{map2}) we divided $p_m$ into probabilities with only even or odd indices. 
For example, one can use the following map
\be \label{maps}
p_{sj +l}\leftrightarrow \mathcal{P}_{jl}
\ee
where $s$ is a fixed integer number greater or equal to $2$, $j=0,1,\ldots, s-1$ and $l$ is a nonnegative integer number. The subadditivity condition~(\ref{subaddit}) in terms of 
$p_m$ has the form
\be 
\fl -\sum_{j=0}^{\infty}\left( \sum_{l=0}^{s-1}p_{sj+l} \right )\ln\left( \sum_{l=0}^{s-1}p_{sj+l} \right ) -\sum_{l=0}^{s-1}\left( \sum_{j=0}^{\infty}p_{sj+l} \right )\ln\left( \sum_{j=0}^{\infty}p_{sj+l} \right )\geq -\sum_{m=0}^{\infty}p_m\ln p_m, 
\ee
which for $s=2$ coincides with~(\ref{ineqentr}). 
For the probabilities expressed in terms of the photon number tomogram of Fock states~(\ref{pm_def2}) we obtain another inequalities for the Laguerre polynomials for arbitrarily $s\geq 2$
\begin{eqnarray}\label{entr_ineq_s}
 -\sum_{j=0}^{\infty}\left( \sum_{l=0}^{s-1}\lambda_{sj+l}(n,x) \right )\ln\left( \sum_{l=0}^{s-1}\lambda_{sj+l}(n,x) \right ) \nonumber \\ \fl
 -\sum_{l=0}^{s-1}\left( \sum_{j=0}^{\infty}\lambda_{sj+l}(n,x) \right )\ln\left( \sum_{j=0}^{\infty}\lambda_{sj+l}(n,x) \right ) 
 +\sum_{m=0}^{\infty}\lambda_m(n,x)\ln \lambda_m(n,x)+x\,\rme^x\geq 0,
\end{eqnarray}
where $\lambda_m(n,x)$ is determined by~(\ref{lambda}).
Let us note that the variation of the mapping impacts the entropies~(\ref{marg_entr}) while the entropy~(\ref{Shannon}) does not change. Hence, one can find such mapping for which the entropic information takes  maximum values.
 
\section{Specific deformation for two mode}
\hspace*{1.5em}
It is possible to extend the notion $f$-oscillator, considered in previous section to the case of two mode. To this end, let us consider the two operators 
\be
\label{2modef}
\hat{A}_1=\hat{a}_1\,f_1(\hat{n}_1,\hat{n}_2),\quad
\hat{A}_2=\hat{a}_2\,f_2(\hat{n}_1,\hat{n}_2),
\ee
where $\hat{a}_1$ and $\hat{a}_2$ are the annihilation operators for the first and second modes and 
$\hat{n}_i=\hat{a}^+_i\hat{a}_i$,\, $i=1,2$. If the functions $f_1$ and $f_2$ satisfy the equation
\be\label{condition}
f_1(n_1,n_2-1)\,f_2(n_1,n_2)=f_1(n_1,n_2)\,f_2(n_1-1,n_2),
\ee
the deformed annihilation operators $\hat{A}_1$ and $\hat{A}_2$ commute, and for this reason one can define the 
two-mode $f$-coherent state $|\alpha_1\,\alpha_2, f_1, f_2\rangle $ by the following equation
\be
\hat{A}_i |\alpha_1\,\alpha_2, f_1, f_2\rangle= \alpha_i |\alpha_1\,\alpha_2, f_1, f_2\rangle.
\ee
Substituting the expansion of the two-mode nonlinear coherent state in the Fock states $|n_1 n_2\rangle$
($\hat{a}^+_i\hat{a}_i|n_1 n_2\rangle = n_i |n_1 n_2\rangle$)
\be
|\alpha_1\,\alpha_2, f_1, f_2\rangle = \sum_{n_1,n_2 =0}^{\infty} C_{n_1, n_2} |n_1 n_2\rangle
\ee
into~(\ref{2modef}), we obtain the recurrence relations for the coefficients $C_{n_1, n_2}$
\be
C_{n_1, n_2}=\frac{\alpha_1}{\sqrt{n_1}\,f_1(n_1,n_2)} C_{n_1-1, n_2},\quad 
C_{n_1, n_2}=\frac{\alpha_2}{\sqrt{n_2}\,f_2(n_1,n_2)} C_{n_1, n_2-1}.
\ee
It worth noting that the latter system of equations is compatible if $f_1$ and $f_2$ satisfy~(\ref{condition}). 

Let us consider the case when the nonlinearity expressed by a one-variable function $f$ depending on the total number of photons $n_{1}+n_{2}$, i.e. $f_1(n_1,n_2)=f_2(n_1,n_2)=f\left( n_{1}+n_{2}\right)$.
It means that the frequency of the oscillator vibrations depends on the
total number of photons or this frequency depends on the sum of the energies
of the two oscillators if one considers not the photon fields but the two
dimensional oscillator. The decomposition of the nonlinear coherent state in the Fock state basis reads

\be\label{series}
|\alpha_1\,\alpha_2,f\rangle = N_f(\alpha_1,\alpha_2)\sum_{n_1,n_2=0}^{\infty}\frac{\alpha_1^{n_1}\alpha_2^{n_2}}{\sqrt{n_1!\,n_2!}\,f(n_1+n_2)!}|n_1\,n_2\rangle,
\ee
where the normalization constant is given by 
\be
N_f(\alpha_1,\alpha_2)=\left (\sum_{n_1,n_2=0}^{\infty}\frac{|\alpha_1|^{2n_1}|\alpha_2|^{2n_2}}{n_1!n_2![f(n_1+n_2)!]^2}  \right )^{-1/2}.
\ee
It can be readily seen from~(\ref{series}) that the nonlinearity creates entanglement. Indeed, 
for the choice $f(n_1+n_2)=1$, the state~(\ref{series}) falls into the separable two-mode coherent state
$|\alpha_1\,\alpha_2,f=1\rangle=|\alpha_1\rangle |\alpha_1\rangle$. In the presence of the nonlinearity function $f$ the state $|\alpha_1\,\alpha_2,f\rangle$ is entangled. To measure entanglement we use the linear entropy defined as
\be
S=1-\mbox{Tr}\hat{\rho}_1^2
\ee
with $\hat{\rho}_1$ being the reduced density matrix of the first subsystem obtained by performing the partial trace of the density operator $\hat{\rho}=|\alpha_1\,\alpha_2,f\rangle \langle\alpha_1\,\alpha_2,f|$ over the second subsystem. The linear entropy ranges from $0$, corresponding to a separable state, and $1$ refers to a maximally entangled state. The linear entropy for the two-mode nonlinear coherent state~(\ref{series}) is
\be \label{linearEntr}
\fl S_f(\alpha_1,\alpha_2)=1-N_f^4(\alpha_1,\alpha_2)\sum_{n,m,p,k=0}^{\infty}\frac{|\alpha_1|^{2(n+p)}|\alpha_2|^{2(k+m)}}{m!\,n!\,k!\,p! f(n+m)!f(p+m)! f(p+k)! f(n+k)!}.
\ee

\begin{figure}[ht]
\begin{center}
\begin{minipage}[h]{0.49\linewidth}
\includegraphics[width=1\linewidth]{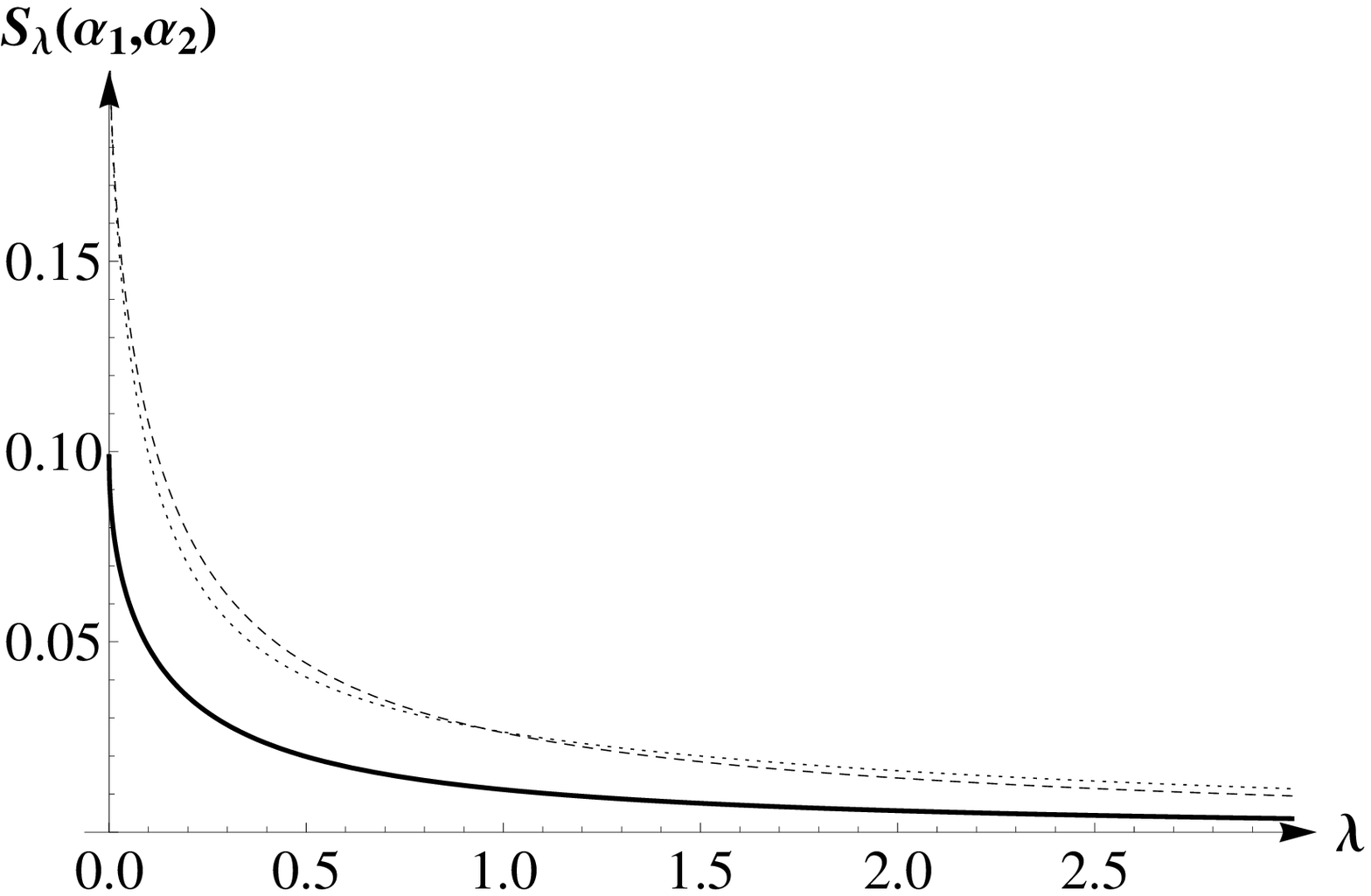}
\caption{The linear entropy $S_{\lambda}(\alpha_1,\alpha_2)$ for $|\alpha_1|=2\,|\alpha_2|=1$ (solid line), $|\alpha_1|=|\alpha_2|=1$ (dashed line), $|\alpha_1|=0.5\,|\alpha_2|=1$ (dotted line) and various values of $\lambda$.}
\label{fig:LinearLambda}
\end{minipage}
\hfill 
\begin{minipage}[h]{0.49\linewidth}
\includegraphics[width=1\linewidth]{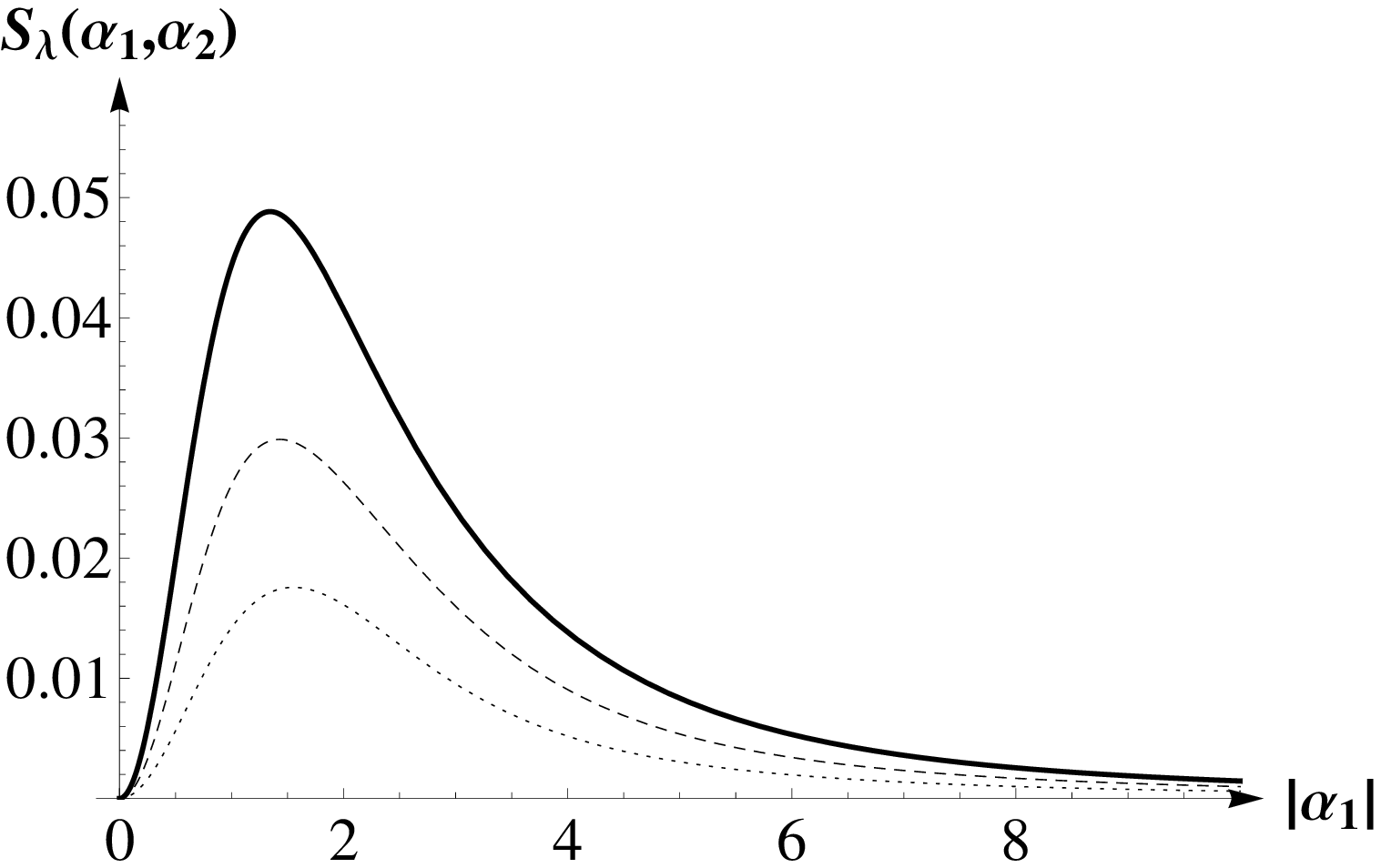}
\caption{The linear entropy $S_{\lambda}(\alpha_1,\alpha_2)$ for $|\alpha_2|=1$ and $\lambda=0.5$ (solid line), $\lambda=1$ (dashed line), $\lambda=2$ (dotted line) and various  values of $|\alpha_1|$.}
\label{fig:LinearAlpha}
\end{minipage}
\end{center}
\end{figure}
Let us note that $S_f(\alpha_1,\alpha_2)$ is symmetric with respect to $\alpha_1$ and $\alpha_2$, i.e. $S_f(\alpha_1,\alpha_2)=S_f(\alpha_2,\alpha_1)$.  
The entanglement properties of $|\alpha_1\,\alpha_2,f\rangle$ are strongly depending on the deformation function. Further we consider deformation function of the form~(\ref{KerrDeform}).
 Figure~\ref{fig:LinearLambda}  depicts the linear entropy $S_{\lambda}(\alpha_1,\alpha_2)$~(\ref{linearEntr}) corresponding to the deformation function~(\ref{KerrDeform}) versus parameter $\lambda$ and different ratios of  $\alpha_1$ and $\alpha_2$. The key role on the behavior of the linear entropy is played by the parameter of
 nonlinearity $\lambda$. The entropy decreases monotonically  with the increase of the value of $\lambda$.  The entropy varies from its maximal value 
\be 
\fl S_{\lambda=0}(\alpha_1,\alpha_2)=1-\left(1+|\alpha_1|^2+|\alpha_1|^2\right)^{-2}\left( 1+
2|\alpha_1|^2+2|\alpha_2|^2 +|\alpha_1|^4+|\alpha_2|^4\right)
\ee 
corresponding to $\lambda=0$ and tends to zero for large values of $\lambda$. Let us note that for $\lambda=0$
the state under consideration has the form
\be
|\alpha_1\,\alpha_2,f_{\lambda=0}\rangle=\left(1+|\alpha_1|^2+|\alpha_2|^2\right)^{-1/2}
\left(|00\rangle+\alpha_1|10\rangle+\alpha_2|01\rangle\right),
\ee
whereas for large values of $\lambda$ the state turns into the usual two-mode coherent state
\be
|\alpha_1\,\alpha_2,f_{\lambda=\infty}\rangle=|\alpha_1\alpha_2\rangle.
\ee
Figure~\ref{fig:LinearAlpha} shows the linear entropy $S_{\lambda}(\alpha_1,\alpha_2)$ as a function of $\alpha_1$
(since $S_{\lambda}(\alpha_1,\alpha_2)$ is symmetric with respect to $\alpha_1$ and $\alpha_2$, we have fixed one of them, for example, $\alpha_2$) for three different values of the parameter $\lambda$. One can see that the maximum value of the linear entropy decreases and shifts towards large values of $|\alpha_1|$ when $\lambda$ increases.
 
The above analysis indicates that the two-mode nonlinear coherent state~(\ref{series}) with the deformation function of the form~(\ref{KerrDeform}) is maximally entangled for close $|\alpha_1|$ and $|\alpha_2|$ and small values of $\lambda$.

\section{Superposition of two-mode nonlinear coherent states}

The Schr\"odinger cat states defined as an even and odd superpositions of coherent states with opposite phases are introduced in~\cite{DodonovMalkinManKo1974} (see also review articles~\cite{Sanders2012,BuzekVidiella-BarrancoKnight1992}). There exist lots of proposal for generating these states (see for example~\cite{dakna1997generating,song1990generation}). The two mode Schr\"odinger cat states  were considered in~\cite{AnsariManko1994}. The Schr\"odinger cat states are found to be a resource for various practical applications. For example, a superposition of coherent states is suggested to be used in interferometers for measuring gravitational waves as alternatives to squeezed states~ \cite{AnsariDiFioreMankoEtAl1994}. In the present work we study even and odd superpositions of the nonlinear coherent states of the form
\be \label{superpos}
|\psi\rangle_f^{\pm}=N_{\pm}(|\alpha\,\alpha,f\rangle \pm|-\alpha\,-\alpha,f\rangle),
\ee
where $|\alpha\,\alpha,f\rangle$ is the two-mode nonlinear coherent state~(\ref{series}). 
The normalization constant is evaluated from the requirement ${}^{\pm}_f\langle \psi|\psi\rangle_f^{\pm}=1$ as follows
\be 
N_{\pm}^{-2}=2\pm 2\,N_f^2(\alpha,\alpha) \sum_{p,q=0}^{\infty}\frac{(-1)^{p+q}|\alpha|^{2p+2q}}{p!\,q![f(p+q)!]^2}.
\ee
The linear entropies of the states $|\psi\rangle_f^{\pm}$ read
\begin{eqnarray}
\label{EntrSuperp}
\fl S^{\pm}_{f}(\alpha)=1-N_{\pm}^4 N_f^4(\alpha,\alpha)\sum_{m,n,k,p=0}^{\infty} \frac{|\alpha|^{2(m+n+k+p)}}{m!\,n!\,k!\,p!}
\nonumber\\
\fl \times\frac{\left( 1\pm (-1)^{m+k}\pm (-1)^{n+k} +(-1)^{m+n} \right) \left( 1\pm (-1)^{m+p}\pm (-1)^{n+p} +(-1)^{m+n} \right)  }{f(m+k)!\,f(n+k)!\,f(m+p)!\,f(n+p)!}
.
\end{eqnarray}

\begin{figure}[ht]
\begin{center}
\begin{minipage}[h]{0.49\linewidth}
\includegraphics[width=1\linewidth]{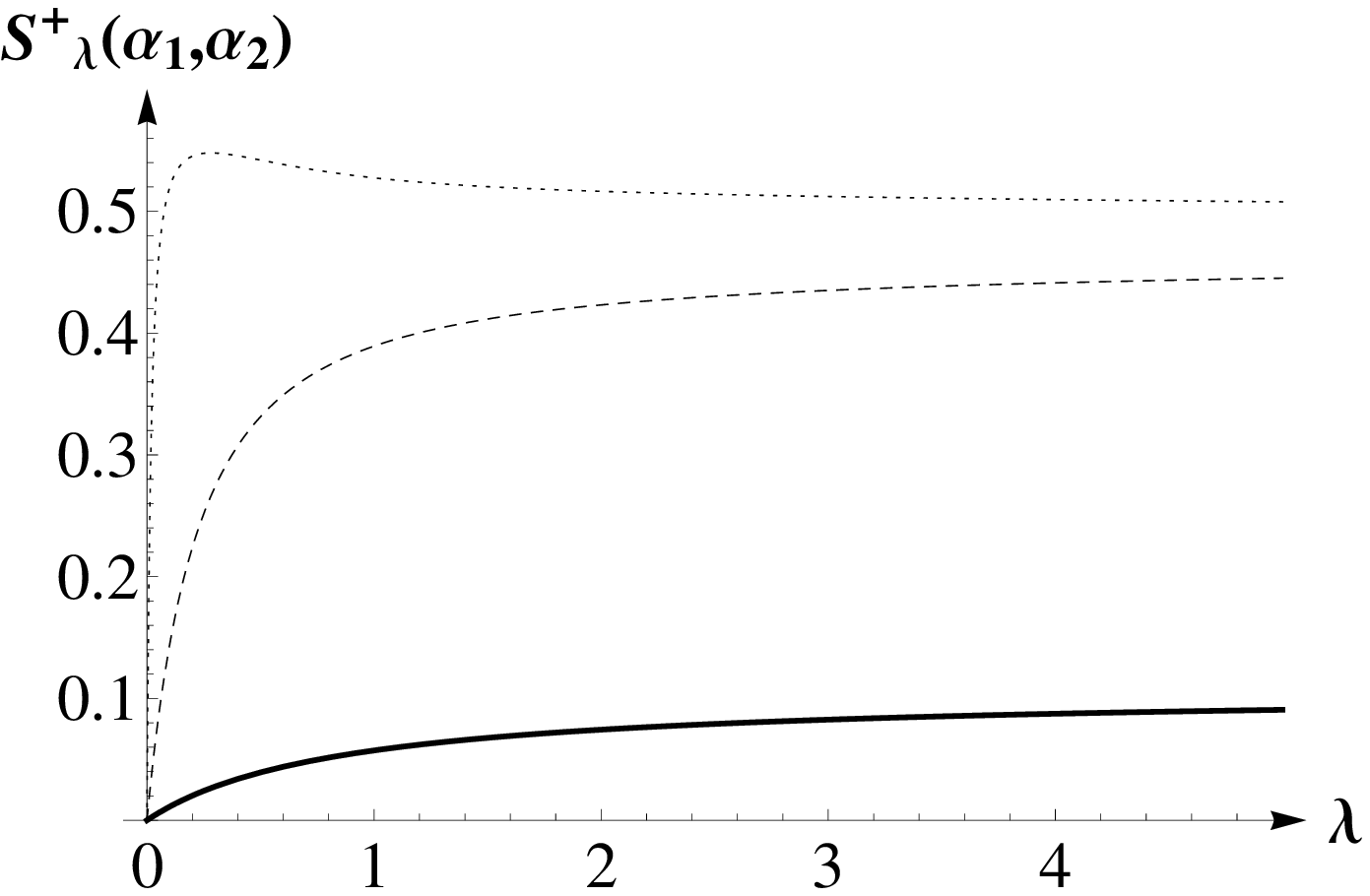}
\caption{The linear entropy $S^+_{\lambda}(\alpha)$ for $|\alpha|=0.5$ (solid line), $|\alpha|=1$ (dashed line),$|\alpha|=2$ (dotted line) and various values of $\lambda$.}
\label{fig:EntrPlus}
\end{minipage}
\hfill 
\begin{minipage}[h]{0.49\linewidth}
\includegraphics[width=1\linewidth]{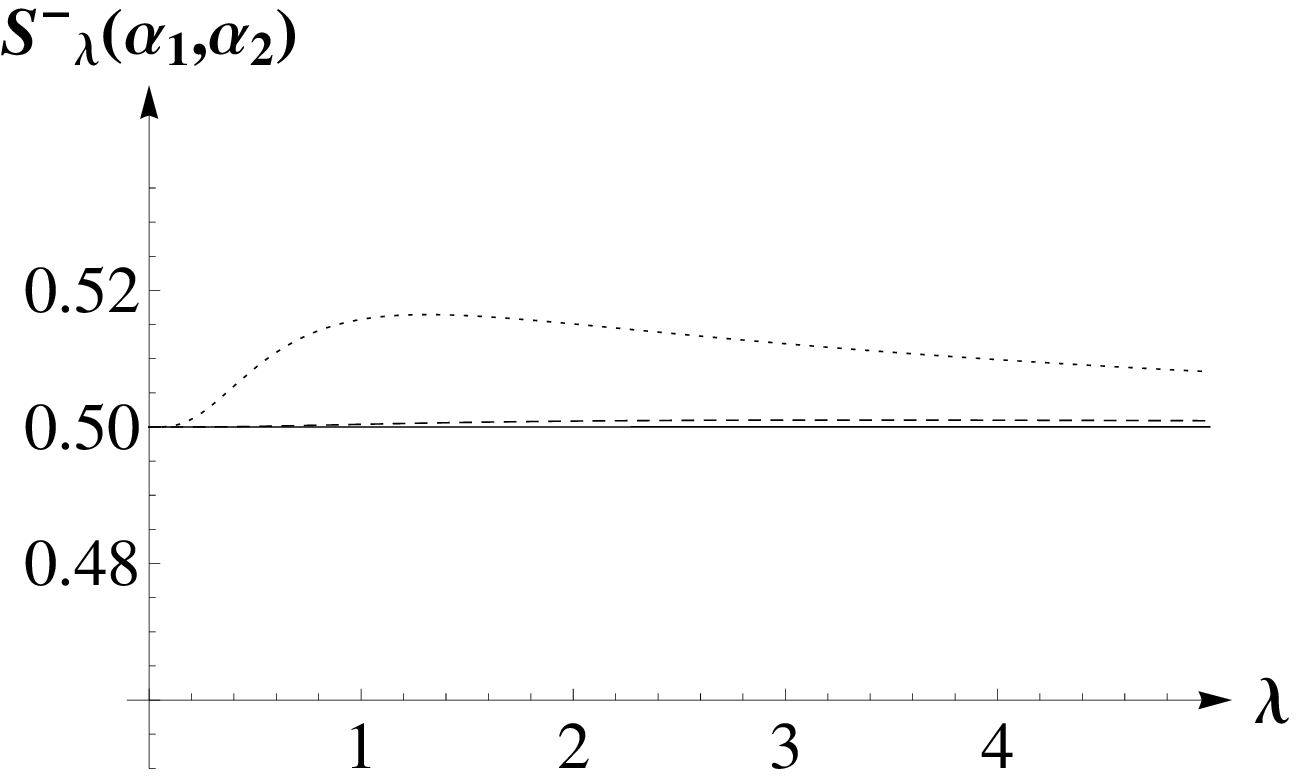}
\caption{The linear entropy $S^-_{\lambda}(\alpha)$ for $|\alpha|=0.5$ (solid line), $|\alpha|=1$ (dashed line),$|\alpha|=2$ (dotted line) and various values of $\lambda$.}
\label{fig:EntrMinus}
\end{minipage}
\end{center}
\end{figure}

The linear entropies~(\ref{EntrSuperp}) for the deformation function~(\ref{KerrDeform}) is plotted in Figures~\ref{fig:EntrPlus} and~\ref{fig:EntrMinus}. For simplicity, we assumed that $\alpha$ is real. One can see that the entanglement properties of the two-mode nonlinear coherent states are considerably different from its superpositions. For $\lambda=0$ the even superposition becomes the product of one-mode vacuum states, i.e. $|\psi\rangle^{+}_{\lambda}=|00\rangle$, whereas the odd superposition is $|\psi\rangle^{-}_{\lambda}=2^{-1/2}(|01\rangle+|10\rangle)$  and corresponding linear entropies are equal to $0$ and $0.5$, respectively. The behaviour of the linear entropies for large values of the parameter $\lambda$ is determined by the fact that the even and odd superpositions of the nonlinear coherent states turn into the usual Schr\"odinger cat states. The linear entropy for the even superposition is
\be
S^+_{\lambda=\infty}(\alpha)=1-0.5\left(1+\rme^{-4|\alpha|^2}\right)^{-2}\left(1+6\,\rme^{-4|\alpha|^2}+\rme^{-8|\alpha|^2} \right)
\ee
and for the odd superposition is $S^-_{\lambda=\infty}(\alpha)=0.5$.

\section{Checking the deformed uncertainty relations}

Pure states satisfy the Heisenberg uncertainty relation~\cite{heisenberg1927anschaulichen} that has the form of inequality for the position and momentum variances. Schr\"odinger~\cite{Schroedinger1930} and Robertson~\cite{Robertson1930} refined this inequality by the correlations of position and momentum. The generalization of the uncertainty relations to the case of mixed states was found in~\cite{dodonov1989invariants}. In~\cite{ManKoMarmoPorzioEtAl2011} the Schr\"odinger-Robertson uncertainty relation was checked experimentally by using homodyne photon detection. Tomographic approach to study uncertainty relations was suggested in~\cite{man2009possible}. It is therefore interesting to study the influence of the considered nonlinearity on the uncertainty relations. 

The Schr\"odinger-Robertson uncertainty relation for the deformed position and momentum operators defined as 

\be
\hat{Q}=\frac{\hat{A}+\hat{A}^{\dagger }}{\sqrt{2}}=\hat{q}f( \hat{a}^{\dagger }\hat{a}),\quad
\hat{P}=\frac{\hat{A}-\hat{A}^{\dagger }}{i\sqrt{2}}=\hat{p}f( \hat{a}^{\dagger }\hat{a}) 
\ee
has the form
\be\label{uncertainty}
\sigma _{QQ}\sigma _{PP}-\sigma^2 _{QP}\geq \frac{1}{4}\left| \langle [\hat{Q},\hat{P}] \rangle \right|^2=
\frac{1}{4}\left| \langle [\hat{A},\hat{A}^{\dagger}] \rangle \right|^2,
\ee
where $\sigma _{QQ}$ and $\sigma _{PP}$ are the variances, $\sigma _{QP}$ is the covariance for $\hat{Q}$ and $\hat{P}$
\be
\fl \sigma _{QQ}=\langle \hat{Q}^2\rangle-\langle \hat{Q}\rangle^2,\quad \sigma _{PP}=\langle \hat{P}^2\rangle-\langle \hat{P}\rangle^2,\quad \sigma _{QP}=\frac{1}{2}\langle \hat{Q}\hat{P}+\hat{P}\hat{Q}\rangle -\langle Q\rangle \langle P\rangle.
\ee
The commutator between the deformed creation and annihilation operators $[\hat{A},\hat{A}^{\dagger}] $ is expressed in terms of the deformation function~(\ref{commut_rel}). 

We consider the $q$-oscillator~\cite{Biedenharn1989,Macfarlane1989}, which is the nonlinear oscillator with the specific exponential dependence of the oscillator frequency on the amplitude of vibrations
\be 
 f_q\left( \hat{a}^{\dagger}\hat{a}\right) =\sqrt{\frac{
\sinh (\lambda \hat{a}^{\dagger}\hat{a})}{\lambda \hat{a}^{\dagger}\hat{a}}}
\ee
with $\lambda$ being a real parameter. The experimental upper bound for the value of the nonlinearity parameter $\lambda$ is given in~\cite{man1995experimental} and is turned out to be small. Therefore,  we consider $\lambda$ to be small  i.e. $\lambda \ll 1$. The uncertainty relation~(\ref{uncertainty}) takes the form
\be \label{Eq:mean_value}
\sigma _{QQ}\sigma _{PP}-\sigma _{QP}^{2}\geq \frac{1}{4}\left( 1+\lambda^2 \left \langle
\hat{n}^2+\hat{n}+\frac{1}{3}
\right \rangle\right),
\ee
where $\hat{n}=\hat{a}^{\dagger}\hat{a}$ is the number of photon.
In the case $\lambda=0$ the latter inequality turns into the uncertainty relation for the usual position and momentum.
But this is not the case in the presence of the nonlinearity the right hand side of~(\ref{Eq:mean_value}) is increased for greater values of $\lambda$. For the ground state the uncertainty relation~(\ref{Eq:mean_value}) reduced to
\be 
\sigma _{QQ}\sigma _{PP}-\sigma _{QP}^{2}\geq \frac{\hbar^2}{4}\left( 1+\frac{\lambda ^{2}}{3}
\right),
\ee
where we have restored the Planck constant
One can interpret the latter inequality  that the presence of the nonlinearity of the $q$-oscillator increases the bound or provides an 'effective Planck constant' value
\be 
\hbar _{eff}^{2}=\hbar ^{2}\left( 1+\frac{\lambda ^{2}}{3}\right).
\ee
In experiments~\cite{BelliniCoelhoFilippovEtAl2012} the commutation relations of the photon creation and annihilation operators were checked by homodyne photon detection
of the photon added and photon subtracted coherent states~\cite
{AgarwalTara1991}. The precision of the bound in this experiment was turned out to be of the order of few percent. If the precision could be at the level of $\lambda ^{2}/3$
with respect to $1$ there would be a possibility to check the influence of the nonlinearity on the value of the effective Planck constant.

The mean value appeared in the right hand side of inequality~(\ref{Eq:mean_value}) can be expressed in terms of  measurable optical tomogram $w\left( X,\theta \right)$
\be
\fl \left \langle
\hat{n}^2+\hat{n}+\frac{1}{3}
\right \rangle=\frac{1}{6}\int X^4\left( w\left(X,0\right)+w\left(X,\frac{\pi}{2}\right)+w\left(X,\frac{\pi}{4}\right)+w\left(X,\frac{3\pi}{4}\right)\right)dX+\frac{1}{12}
\ee
Thus we propose to check the contribution of the $q-$nonlinearity to the
bound of the uncertainty relation equivalent to increasing the value of
the effective Planck constant. Other types of nonlinearity, e.g. proposed in~ 
\cite{BastosBertolamiDiasEtAl2012} also can be checked in the same experiments.

\section{Conclusions}
To conclude, we point out the main results of this paper.
We reviewed the properties of the quantum nonlinear oscillators
vibrating with frequencies depending on the amplitude of vibrations. We
reviewed the tomographic description of quantum states using the optical,
symplectic, as well as photon number tomographies. We used the known subadditivity condition
 for joint probability distributions as well as the explicit form of the 
photon number tomography for the Fock states to obtain a new inequality for the associated 
Laguerre polynomials. We studied
the properties of the nonlinear $f$-oscillators in the framework of the
tomographic pictures of quantum mechanics. The quantum tomograms for the nonlinear coherent states are obtained in explicit form. The contribution of small nonlinearity vibrations to the uncertainty
relations bound was obtained for the $q$-oscillators. The possibility to
study the uncertainty relation in experiments with homodyne detection of
photonic states is considered in~\cite{MankoMarmoZaccariaEtAl2005}. The formalism of the $f$-oscillators can be applied also to study different phenomena in solid state statistics and gravitation physics. 
\ack 
V. I. thanks the Russian Science Foundation for financial support under Project No. 16-11-00084.
G.M. would like to acknowledge the grant 'Santander-UC3M Excellence Chairs 2016/2017'.

\section*{References}
\bibliography{literatura}
\end{document}